\documentclass[aps,prl,floatfix,twocolumn,groupedaddress]{revtex4}
\usepackage{epstopdf}
\usepackage{times}
\usepackage{amssymb,amsmath}
\usepackage{array}
\usepackage{multirow}
\usepackage{color}
\usepackage{setspace}
\usepackage{changes}
	\definechangesauthor[name={Per cusse}, color=orange]{per}

\usepackage{hyperref}
\hypersetup{hypertex=true,
    colorlinks=true,
    linkcolor=blue,
    anchorcolor=blue,
    citecolor=blue}

\usepackage{braket}
\usepackage{booktabs}
\usepackage{color}
\usepackage{threeparttable}
\usepackage{multirow}
\usepackage{mathtools}
\usepackage{graphicx}

\renewcommand\keywords[1]{\textbf{Keywords}: #1}

\makeatletter

\newcommand{\mum}{\,\mu\mbox{m}}

\newcommand{\ps}{\,\mbox{ps}}

\begin{document}

\title[]{Ten-channel Hong-Ou-Mandel interference between independent optical combs}

\author{Wenhan Yan$^{1,\dagger}$, Yang Hu$^{1,\dagger}$, Yifeng Du$^{1}$, Kai Wang$^{1}$, Yan-Qing Lu$^{1}$, Shining Zhu$^{1}$, Xiao-Song Ma$^{1,2,3,\ast}$}

\affiliation{
$^1$~National Laboratory of Solid-State Microstructures, Collaborative Innovation Center of Advanced Microstructures, School of Physics, Nanjing University, Nanjing 210093, China\\
$^2$~Synergetic Innovation Center of Quantum Information and Quantum Physics, University of Science and Technology of China, Hefei, 230026, China\\
$^3$~Hefei National Laboratory, Hefei 230088, China\\ 
$^{\dagger}$These authors contributed equally to this work\\
$^{\ast}$Corresponding author: xiaosong.ma@nju.edu.cn
}
\date{\today}

\begin{abstract}	

Dissipative Kerr soliton (DKS) frequency comb exhibits broad and narrow-linewidth frequency modes, which make it suitable for quantum communication. However, scalable quantum network based on multiple independent combs is still a challenge due to their fabrication-induced frequency mismatches. This limitation becomes critical in measurement-device-independent quantum key distribution, which requires high visibility of Hong-Ou-Mandel interference between multiple frequency channels. Here, we experimentally demonstrate two independent DKS combs with ten spectrally aligned lines without any frequency locking system. The visibility for individual comb-line pairs reaches up to $46.72 \pm 0.63\%$ via precision frequency translation, establishing a foundation for deploying DKS combs in multi-user quantum networks.
	
\noindent

\end{abstract}

\maketitle
\twocolumngrid
\section{Introduction}

In the quantum key distribution (QKD)~\cite{cryptography1984public,gisin2002quantum,scarani2009security,xu2020secure}, the single-frequency lasers are widely used as light source to generate the weak coherent states (WCS). Going beyond the point-to-point QKD, the large-scale quantum networks with many users and simultaneous communication links require multi-frequency light sources and wavelength-division multiplexing (WDM) technology to improve communication performance and efficiency. As a direct solution for integrated QKD transceiver chips~\cite{sibson2017chip}, combining multiple lasers with different wavelengths is experimentally feasible, however, it has substantial challenges in scalability. 

In recent years, the development of manufacturing processes of compact chip-scale micro-ring resonators (MRRs)~\cite{kippenberg2011microresonator} provides a more efficient way to acquire highly coherence multi-frequencies light such as dissipative Kerr soliton comb~\cite{herr2016dissipative} , of which the formation is governed by a double-balance of nonlinearity and dispersion, as well as dissipation and gain. The distinctive compact~\cite{stern2018battery}, low-noise~\cite{liang2015high}, low-power comsumption~\cite{stern2018battery,liu2018ultralow} frequency comb has emerged as a promising light source for a range of technologies~\cite{chembo2016kerr}, including optical clocks~\cite{ghelfi2014fully}, optical frequency synthesis~\cite{spencer2018optical,briles2018interlocking}, classical communication~\cite{wang2020quantum,marin2017microresonator,pfeifle2014coherent,geng2022coherent}, and sensing~\cite{ndagano2022quantum,wang2024nanometric}. The advent of the optical frequency comb has enabled the provision of a diverse and abundant single-frequency resource with highly coherent property. The dissipative Kerr soliton can generate a single optical comb with a high signal-to-noise ratio (SNR) across a broad band (150 nm), encompassing both C and L bands~\cite{li2017stably,pfeiffer2017octave,okawachi2014bandwidth,song2024octave}. The combination of an optical comb with WDM technology represents a crucial step toward the integration and scalability of multi-user quantum networks~\cite{yan2025measurement}. The indistinguishability of independent light sources belonging to different users is essential for the measurement-device-independent QKD (MDI QKD)~\cite{braunstein2012side,lo2012measurement,zheng2021heterogeneously}, which is based on the two-photon Hong-Ou-Mandel (HOM) interference~\cite{hong1987measurement,kim2020hong}. Therefore, to obtain high-visibility HOM interference between frequency lines generated from different independent micro-ring resonators is highly desirable.

Despite progress in frequency locking field such as the f-2f ~\cite{brasch2017self,newman2019architecture} and atomic reference schemes~\cite{Papp:14}, which can achieve frequency stability of  $9.5 \times 10^{-14}$@1s~\cite{niu2024atom}, it remains challenging to attain a high degree of homogeneity across all comb lines of two independent DKS combs. This challenge arises from the differences between independent MRRs due to fabrications. Considering the free spectral range (FSR) of the micro-ring in the first order approximation of dispersion: $FSR(v) = c/n_g \times L$ , even minor discrepancies in the length of the cavity (L) or imperfections in the waveguide can result in a deviation of the FSR.

In this work, we generate two independent DKS combs from different dispersion-engineered silicon-nitride microring resonators simultaneously, and ten high SNR comb-lines are successfully obtained with WDM. For each DKS comb generation system, the auxiliary laser–assisted intracavity thermal balance technique~\cite{lu2019deterministic,zhou2019soliton,zhang2019sub} and the temperature control have been employed to enhance the overall stability, sustaining operation for over 8 hours with two free-running lasers~\cite{geng2020enhancing}. The integration of micro-ring heaters on the chip and the single-sideband (SSB) modulator have been employed to eliminate the frequency differences of the combs at several free spectral ranges (FSRs) away from the pump. Additionally, high-speed modulators chop continuous light into ultra-narrow pulses with a width of about 0.4 ns. Finally, we demonstrate the HOM interference between ten independent comb lines and obtain high visibility between them. The results show the great potential of DKS combs as a quantum communication network light sources.

\section{Experimental Setup}
\begin{figure*}[htbp]
\begin{center}
    \includegraphics[width=1\textwidth]{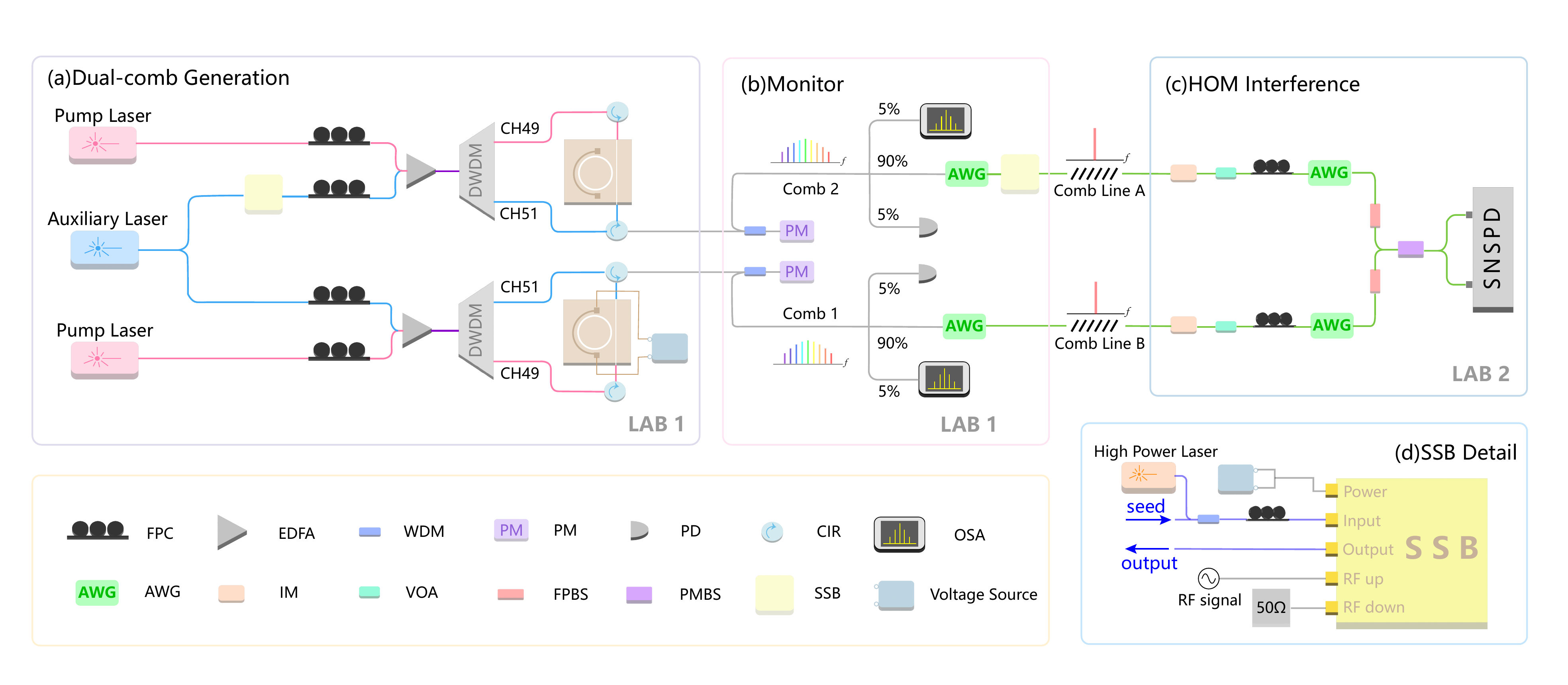}
    \caption{\label{Fig1}  The experimental setup of the HOM interference measurement system. \textbf{(a)}. Two independent micro-ring resonators are fabricated on a Silicon Nitride chip and generate DKS combs simultaneously. \textbf{(b)}. Two sets of ten comb-lines are filtered using dense wavelength de-multiplexing (DWDM) fiber filters and circulators. \textbf{(c)}. Each comb-line possesses an encoder consisting of an intensity modulator (IM) that generates short time-domain pulse. The following variable optical attenuator (VOA) is used to attenuate the pulses to the single-photon level, and the two-photon HOM interference is measured by superconducting nanowire single-photon detectors (SNSPDs). \textbf{(d)}. Setup of the single-sideband (SSB) modulator section : a voltage source, which is constantly loaded on 12V , is used as the power supply of the modulator. In order to achieve high carrier extinction, we use a single frequency laser with a central wavelength of CH19 as an auxiliary light source to lock the working point of the modulator. The microwave source is connected to one of the RF ports (RF-up) to load the frequency-shifted signal and the other terminal (RF-down) is connected to a 50 $\Omega$ load to prevent reflection. After sideband modulation, the signal light is divided by a CH19 WDM filter to isolate the interference of the auxiliary light. The genration and monitor part of DKS combs have been set in Lab 1, and the HOM interference part is set in Lab 2.  FPC, fiber polarization controller; EDFA, erbium-doped optical amplifier; WDM, 200 $GHz$ waveLength division multiplexing filter; PM, powermeter; PD, power detector; CIR, three-port circulator; OSA, optical spectral analyzer; AWG, arrayed waveguide grating; IM, intensity modulator; VOA, variable optical attenuator; FPBS, fiber polarization beam splitter; PMBS, polariaztion-maintaining beam splitter; SSB, single-sideband modulator; SNSPD, superconducting nanowire single-photon detector.} 
\end{center}
\end{figure*}

The experimental setup is illustrated in Fig. \ref{Fig1}. The optical comb light source is based on an integrated silicon nitride platform (Ligentec AN800) comprising two micro-resonator devices with identical cavity design parameters. Two tunable semiconductor lasers are used as main pump sources, and a single frequency laser with narrow linewidth serves as the auxiliary laser applied to heat the micro-resonator from the opposing direction, which is based on the dual-driven technique ~\cite{zhou2019soliton,zhang2019sub,geng2020enhancing} for stably generating the DKS comb. The light polarization on the chip is set to the transverse electric (TE00) mode, which exhibits higher loaded quality factors and extinction ratios. The employment of WDM fiber filters combines light from different pump channels into erbium-doped optical amplifiers (EDFAs), which are configured to enhance the light power to compensate for optical path and coupling losses. Two cascaded three-port circulators are used to implement a dual-driven soliton access system. The power meters (PMs) and photodetectors (PDs) are used to record changes in the power spectrum of the pump and the soliton comb, respectively, which help determine the soliton-step position. Simultaneously, the optical spectrum analyzer records the frequency spectrum of the corresponding soliton states.

The implementation of dense wavelength de-multiplexing (DWDM) and a 32-channel arrayed waveguide grating (AWG) filters the residual pump light, thereby isolating the micro-cavity frequency comb with a 100 GHz frequency spacing. Here, we use high-precision frequency translation with single-sideband modulation, combined with the on-chip micro-heater, to align the frequencies between different DKS comb lines. 

At this stage, the two overlapping spectra signal photons are directed to the subsequent encoding apparatus. A commercial intensity modulator splits continuous-wave light from a single DKS comb line into a series of narrow pulses. The electrical signals driving the intensity modulators (IMs) are generated by an arbitrary waveform generator. A variable optical attenuator (VOA) attenuates the pulses to the single-photon level for two-photon Hong-Ou-Mandel (HOM) interference measurements. 

\section{The generation of Dissipative Kerr solitons}

\begin{figure*}[htbp]
	\begin{center}
		\includegraphics[width=0.8\textwidth]{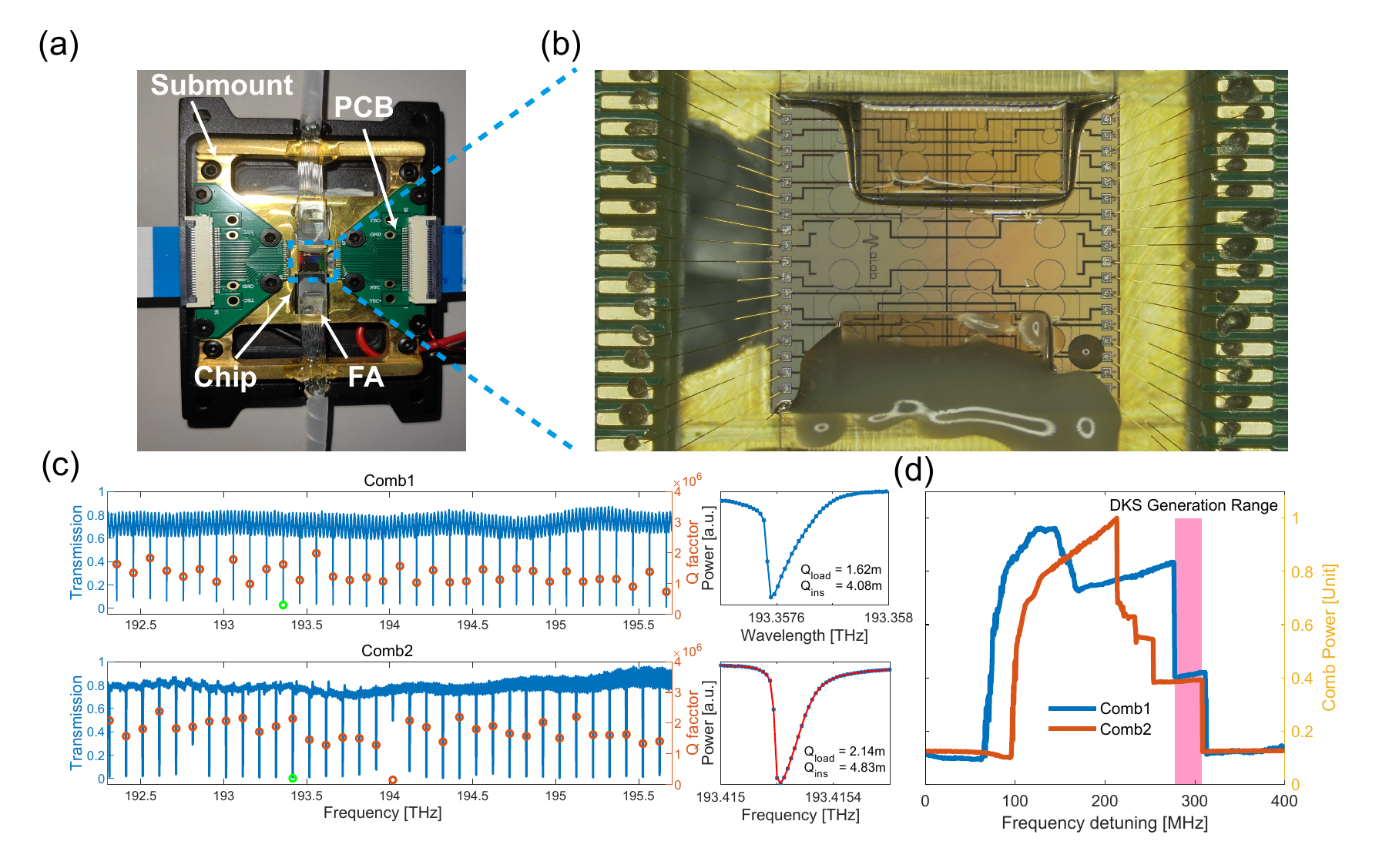}
		\caption{\label{Fig2}  \textbf{(a)}. Light coupling and packaging using 32-Channel ultra-high numerical aperture (UHNA7) fiber arrays. \textbf{(b)}. Silicon nitride chip under optical microscope. \textbf{(c)}. The transmission spectrum of two independent microcavities. The red circles represent the loaded quality factors derived from each near-cold resonances and the green hollow circles mean the location of the main pump laser, respectively. The middle illustration represents the intrinsic quality factor data for the main pump of two different micro-cavity. For the Cavity 1(Cavity 2), the intrinsic quality factor is about 4.1(4.8) million. \textbf{(d)}. The simultaneous generation of two DKS combs. The comb power record of two DKS combs with the high-precision frequency detuning of the pump light. Two DKS combs can be produced when the frequency of the main pump is reaching the pink region, respectively.} 
	\end{center}
\end{figure*}

We use optical and electrical packaging (see Fig.  \ref{Fig2}\textbf{(a)}) to accommodate the silicon nitride chip, which is located at the center of the chip holder, features an inverted taper design to enhance the coupling of light into and out from the chip. It is then coupled to the ultra-high numerical aperture (UHNA7) fiber array. The coupling point between the chip and the optical fiber array is fixed with UV-curable glue (see the dark part of the image), which ensures the long-term stability of the system coupling after high temperature aging. The printed circuit board (PCB) is designed to extend the electrical port for the on-chip micro-ring heaters using gold and soft wires, with the final exportation located at the other end. The submount provides support for the chip temperature control system, enabling the transfer of heat generated by the operational chip through the oxygen-free copper to the cooling sheet for thermal feedback. 

A self-made  wavelength calibrated system based on an asymmetric Mach–Zehnder interferometer (AMZI) is used to record over one hundred resonant frequencies of the TE00 mode, from which the cavity group velocity dispersion (GVD) is extracted using the equation $\omega_\mu = \omega_0 +\sum_{j}D_j\mu^j/j!$.
The total quality factor ($Q_{tot}$) and extinction ratio (ER) are given by:
\begin{equation}
Q_{tot} = \frac{{\pi}n_gL\sqrt{ra}}{\lambda_{res}(1-ra)}  , 
\end{equation}
\begin{equation}
ER = 10\lg\frac{(r-a)^2(1+ra)^2}{(r+a)^2(1-ra)^2}  ,
\end{equation}
Based on the aforementioned equations~\cite{bogaerts2012silicon,zhao2021integrated}, the intrinsic quality factors of Cavity 1 and Cavity 2 are determined to be about 4.1$\times10^6$ and 4.8$\times10^6$ , respectively. COMSOL simulations are employed to estimate the effective mode area of the waveguide, which is found to be 1.26$\mum^2$. Consequently, the parametric oscillation~\cite{kippenberg2004kerr,li2012low,briles2020generating,savchenkov2004low} thresholds of the two microcavities are calculated as about 2.66 $mW$ for Cavity 1 and 2.41 $mW$ for Cavity 2, respectively. The measured fiber-chip-fiber insertion losses are 5.375 dB/facet for Cavity 1 and 4.880 dB/facet for Cavity 2, including propagation losses in the on-chip waveguides. The pump power for both cavities is amplified to approximately 23 dBm. The on-chip power of Cavity 1 is approximately 17 dBm, while the on-chip power of Cavity 2 is approximately 16.5 dBm. With the fine-detuning of the main laser's frequency, we can access the soliton platform at a relative slower rate. And a smooth single-soliton state can be reached.

\section{Alignment of the frequency}

To achieve high visibility in independent Hong-Ou-Mandel (HOM) interference for two independent DKS combs, it is necessary to eliminate the frequency differences between their comb lines. In order to simplify the experimental setup of the detection side, we put the frequency compensation part in the light source generation part. The frequencies of the comb lines can be expressed as $f_{\mu} = f_p \pm \mu*FSR$, where $\mu$ is the relative mode number with respect to the pump and $f_p$ is the frequency of the pump. Due to the thermo-optical effect, electrically driving the integrated heater changes the refractive index of the waveguide, thereby adjusting the spectral positions of the cavity resonances. Similarly, an optical signal spectrum containing modulation frequency information is generated through an electro-optical phase modulation process. Considering that the combination of a single-sideband modulator and a micro-heater is used for fine adjustment of comb-line frequencies, the frequency difference ${\Delta}f_{\mu}$ between different combs can be expressed as 
\begin{equation}
{\Delta}f_{\mu} = (f_{p_1}-f_{p_2}) + \mu*{\Delta}FSR + f_{SSB} + f_{heater}
\end{equation}
,where ${\Delta}FSR$ is the free spectral range difference with different cavities, $f_{p_1}(f_{p_2})$ is the frequency of the pump of comb1(comb2), $f_{SSB}$ is the microwave signal frequency applied to the single-sideband modulator and the $f_{heater}$ is the frequency shift component by the micro-heater.

Once the robust soliton state of the two devices has been obtained, the crucial challenge is to control the pump laser in such a way that the same soliton states of the two micro-rings are accessed simultaneously. The slight difference in cavity length between the two micro-rings results in a difference in free spectral range (FSR), with $FSR_2$ = 100.54GHz being slightly larger than $FSR_1$ = 100.41GHz. This naturally leads to a separation of the natural frequencies between the two frequency combs. To align the selected signal cavity mode (from CH38 to CH44), a voltage of approximately 7.85 V is applied to the integrated micro-heater on Cavity 2, utilizing the thermo-optical effect to induce a red shift in the entire Cavity 2 spectrum. The adjustment accuracy is 0.01V, and the response is 2.26$pm/mW$, with a maximum theoretical adjustment range of approximately 50 $GHz$. 

Additionally, a single-sideband (SSB) modulator is loaded on one of the optical paths of the auxiliary pump, allowing the simultaneous access of soliton platforms of two cavities. Given the employment of two independent main pump sources, it is imperative to consider the initial pump frequency differences for the purpose of compensation. Conversely, the relative frequency difference of the main pump position can be utilized for fine-tuning the comb lines, as the Fig. \ref{Fig2} \textbf{(d)} shows that the tuning length is allowed to exceed 30 $MHz$ while maintain the soliton states. The system's stability has been demonstrated to exceed 8 hours under these conditions(as shown in Fig. \ref{Figs2}).

\section{HOM interference results}

\begin{figure*}[htbp]
\begin{center}
    \includegraphics[width=1\textwidth]{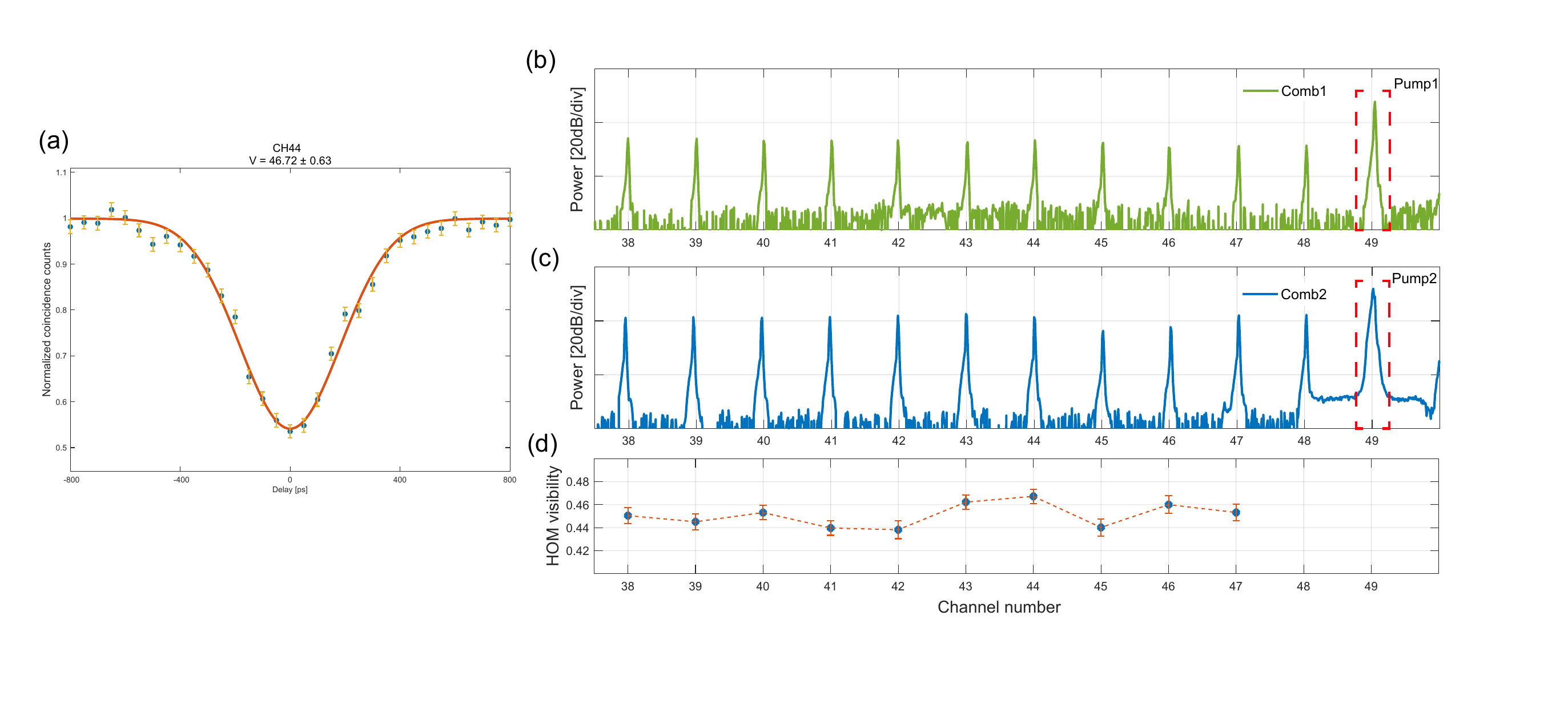}
    \caption{\label{Fig3} HOM interference visibility diagram of signal light after alignment. \textbf{(a)}. the two photon HOM interference data of the CH44, and the visibility is estimated to be $46.72 \pm 0.63\%$.  \textbf{(b)}.and\textbf{(c)}. The aligned 10 comb lines spectrum of the two independent DKS combs \textbf{(d)}.  the HOM interference visibility data of 10 alignment comb lines from CH38 to CH47.} 
\end{center}
\end{figure*}

The DKS comb is filtered through the cascaded 32-channel arrayed waveguide grating filters with a 100 $GHz$ bandwidth, resulting in a single-frequency light with a signal-to-noise ratio exceeding 50dB (as the Fig. \ref{Figs3} \textbf{(f)} shows). The single-frequency comb lines are transmitted to the detection setup in another laboratory (Lab 2) through a single-mode optical fiber cable of about 600 meters. An intensity modulator is used to prepare the WCS pulses. A polarization controller adjusts the polarization of the signal light before it enters the intensity modulator. The continuous light is chopped to pulse series by IMs, on which the electrical signal with repetition rate of 250 MHz and width of about 400 ps is loaded.

The light pulses are projected into the Hong-Ou-Mandel(HOM) interferometer via a combination of a fiber polarization beam splitter and a fiber polarization-preserving beam splitter, which maintains the polarization of the two photons. The output photons are detected by superconducting nanowire single-photon detectors (SNSPDs) and a field-programmable gate array (FPGA)-based coincidence logic unit records and analyses the photon detection signals. The coincidence count probability $P_{coin}(\tau)$ is a function of delay time $\tau$ between pulses of two users :
\begin{equation}
P_{coin}(\tau) = \frac{1}{2} - \frac{1}{4}V{e^{-\frac{{\tau}^2}{2{\sigma}^2}}}cos(\tau{\Delta}\omega) ,
\end{equation}
where $V$ represents the fringe visibility of HOM interference, $\sigma$ is the full width at half maximum of the wave packet, and ${\Delta}\omega$ is the frequency difference of the two interference photons~\cite{rarity2005non}. As the Fig. \ref{Fig3} shows, the multichannel HOM interference has been demonstrated from the CH38 to CH47. The spectrums of two soliton states are recorded separately by a spectral analyzer.In order to keep the pump mode (as shown in the dotted box of the Fig. \ref{Fig3} \textbf{(b)} and \textbf{(c)}) always in the flat position of the filter bandwidth when adjusting the frequency within the range of a FSR, the 200 $GHz$ WDM is used as the post filter. After compensating the frequencies of different pairs of combs respectively, we performed a  time delay scan from -800 $\ps$ to 800 $\ps$ for the step of 50 $\ps$. The experimental results are shown in Fig. \ref{Fig3} \textbf{(c)}  , where the visibilities of HOM interference are distributed from $43.1 \pm 0.6\%$ to $46.7\pm 0.6\%$. To evaluate the stability, the HOM interference of every comb-teeth pair is measured five times.
\section{Conclusion}
In this work, we demonstrate the frequency interference between different independent DKS combs using high-precision frequency translation and the on-chip thermo-optical tuning. We have measured HOM interference visibilities of 10 pairs of single comb lines. And the average visibility ranges from $43.14\%$ to $46.72\%$, demonstrating that the independent DKS optical combs exhibit good indistinguishability comparable to that measurement results using single-frequency lasers. Without using any frequency locking system, the soliton system platform remained stable for 8 hours with the free-running lasers. It is anticipated that a longer-term, more stable multi-frequency broadband light source will be obtained if the pump laser is locked to the ultra-stable cavity. The findings of our research provide a possibility of a scalable and integrated platform for the realization of multi-user quantum networks.

\section*{Funding Sources}
This research was supported by the National Key Research and Development Program of China (Grants No. 2022YFE0137000 ), Natural Science Foundation of Jiangsu Province (Grants Nos. BK20240006,BK20233001), the Leading-Edge Technology Program of Jiangsu Natural Science Foundation (Grant No. BK20192001), the Fundamental Research Funds for the Central Universities, and the Innovation Program for Quantum Science and Technology (Grants Nos. 2021ZD0300700 and 2021ZD0301500), Supported by the Fundamental Research Funds for the Central Universities (Grants Nos. 2024300324).\noindent

Note : During the finalization of this project, we become aware of a related work is published ~\cite{Massively}.

\noindent
\bibliographystyle{naturemag}
\bibliography{DKS_ref1}

\clearpage
\newpage
\onecolumngrid 
\section{Supplementary information}

\subsection{The generation of the DKS combs}\label{sec:one}

\begin{figure*}[htbp]
	\begin{center}
		\includegraphics[width=1\textwidth]{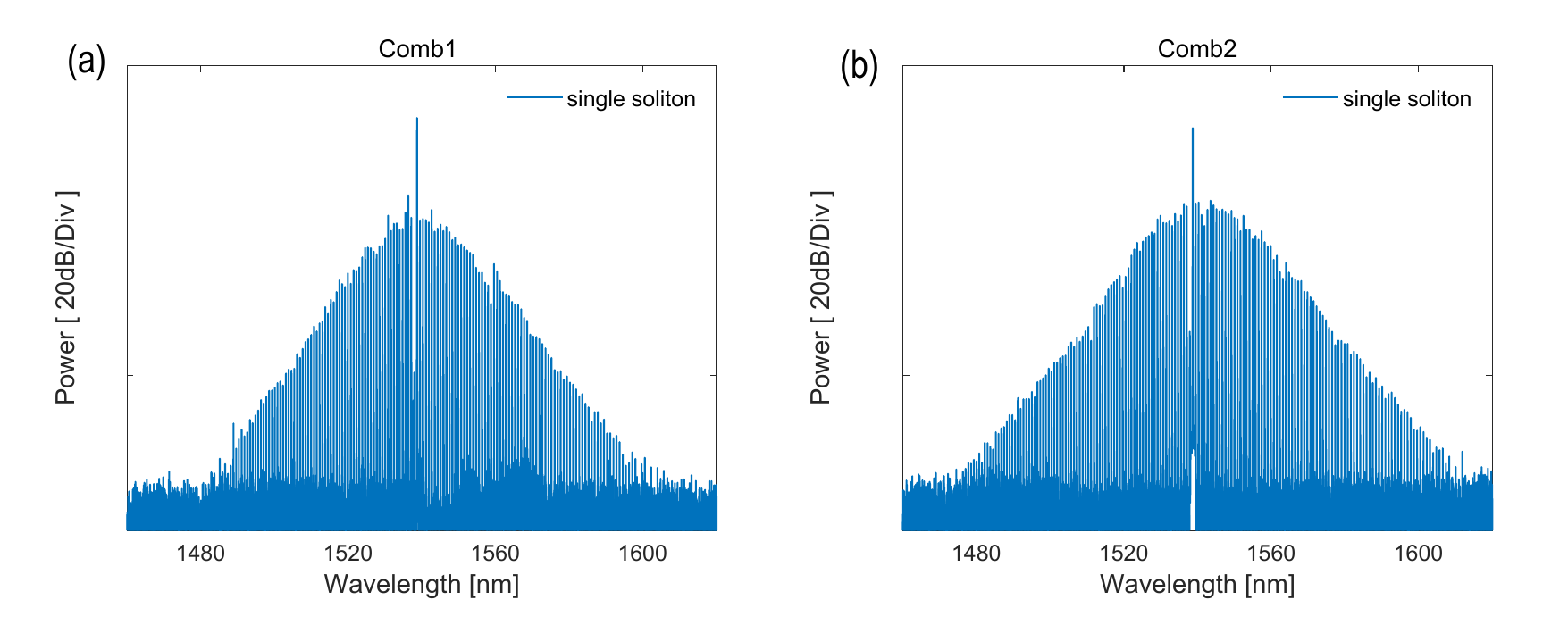}
		\caption{\label{Figs1} From \textbf{(a)} to \textbf{(b)} are the measurement single soliton frequency sprctrums of two independent DKS combs. } 
	\end{center}
\end{figure*}

\begin{figure*}[htbp]
	\begin{center}
		\begin{minipage}{0.45\textwidth}
			\centering
			\includegraphics[width=1\textwidth]{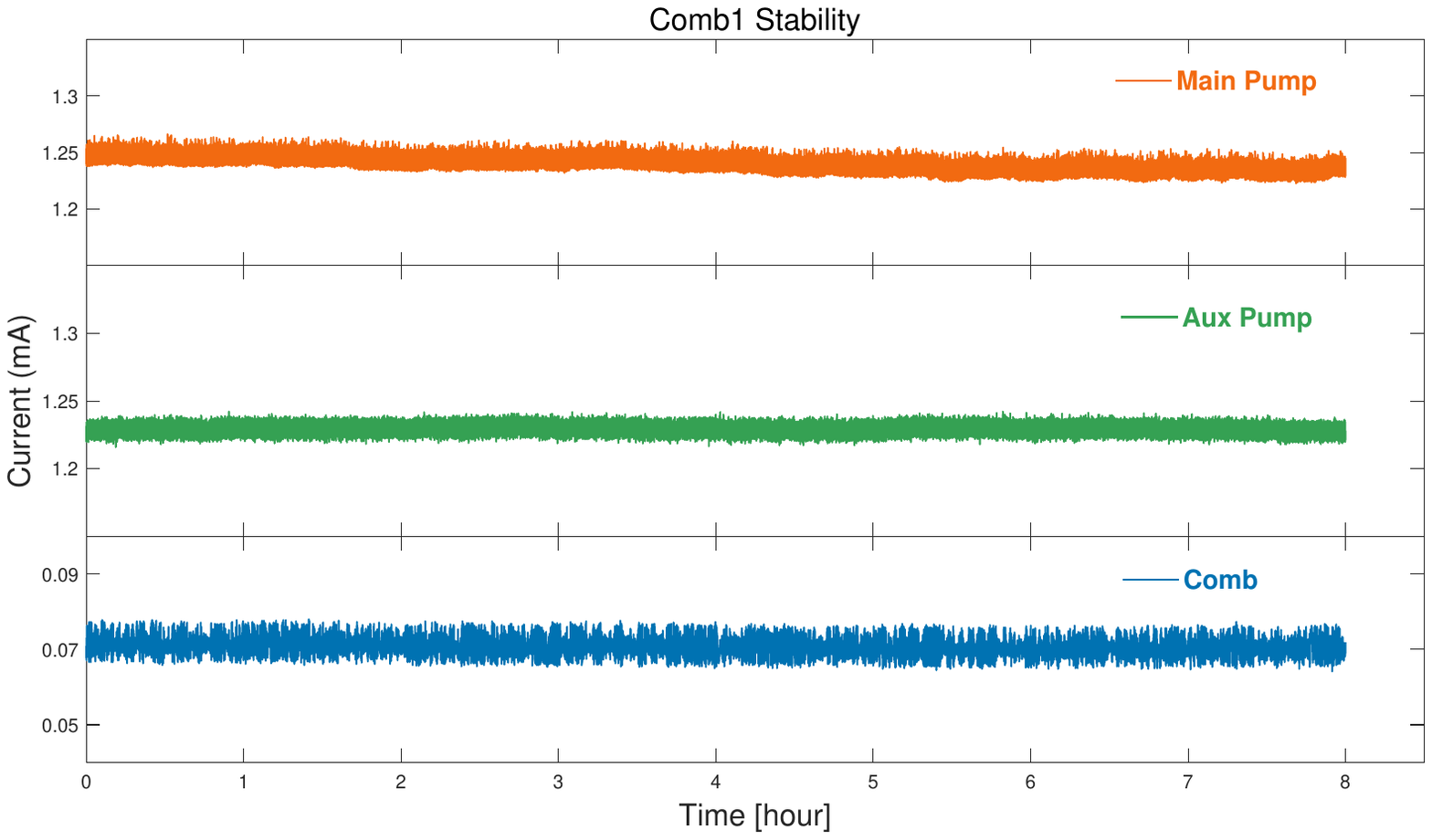}
		\end{minipage}\hfill
		\begin{minipage}{0.45\textwidth}
			\centering
			\includegraphics[width=1\textwidth]{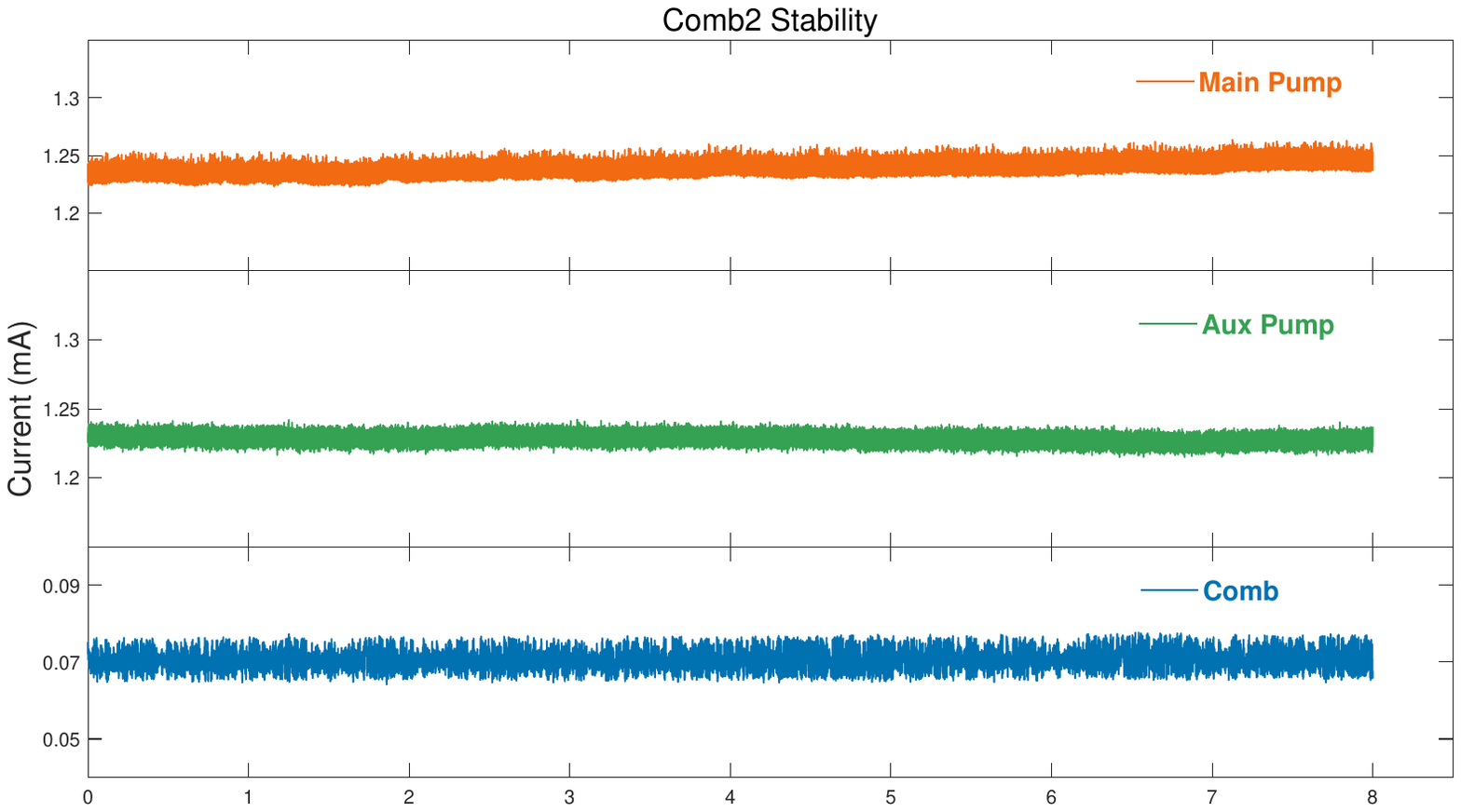}
		\end{minipage}
		\caption{\label{Figs2} The above figure shows the power stability of the aux pump, main pump and the comb power of Comb1 and Comb2 over a two-hour period, respectively.}
		
	\end{center}
\end{figure*}

Stable pumping of a high-Q micro-resonator with narrow linewidth needs to overcome environmental perturbations. And the existence of the sizable cavity thermal nonlinearity when accessing the DKS parameter space makes it easier to be out of resonance. The heat flow of the main pump can be largely balanced out with the approppriate power and frequency of the auxiliary pump, keeping the cavity temperature and all cavity resonances approximately unchanged. This allows the pump laser to be stably tuned across the entire micro-resonance with minimized thermal behavior. Based on the dual-driven system , the overall stability of the DKS combs generation system has been improved even stay in the single soliton state, which is shown in the Fig. ~\ref{Figs1}. With a frequency detuning accuracy of about 0.5 $MHz$ minimum step size, we can slowly access the soliton regions on the red detuned side of two resonators simultaneously and maintain a stable time of more than 8 hours (see Fig. ~\ref{Figs2}).
 
\subsection{The results of 32 channel arrayed waveguide grating (AWG) filter}\label{sec:three}

\begin{figure*}[htbp]
	\begin{center}
		\includegraphics[width=0.7\textwidth]{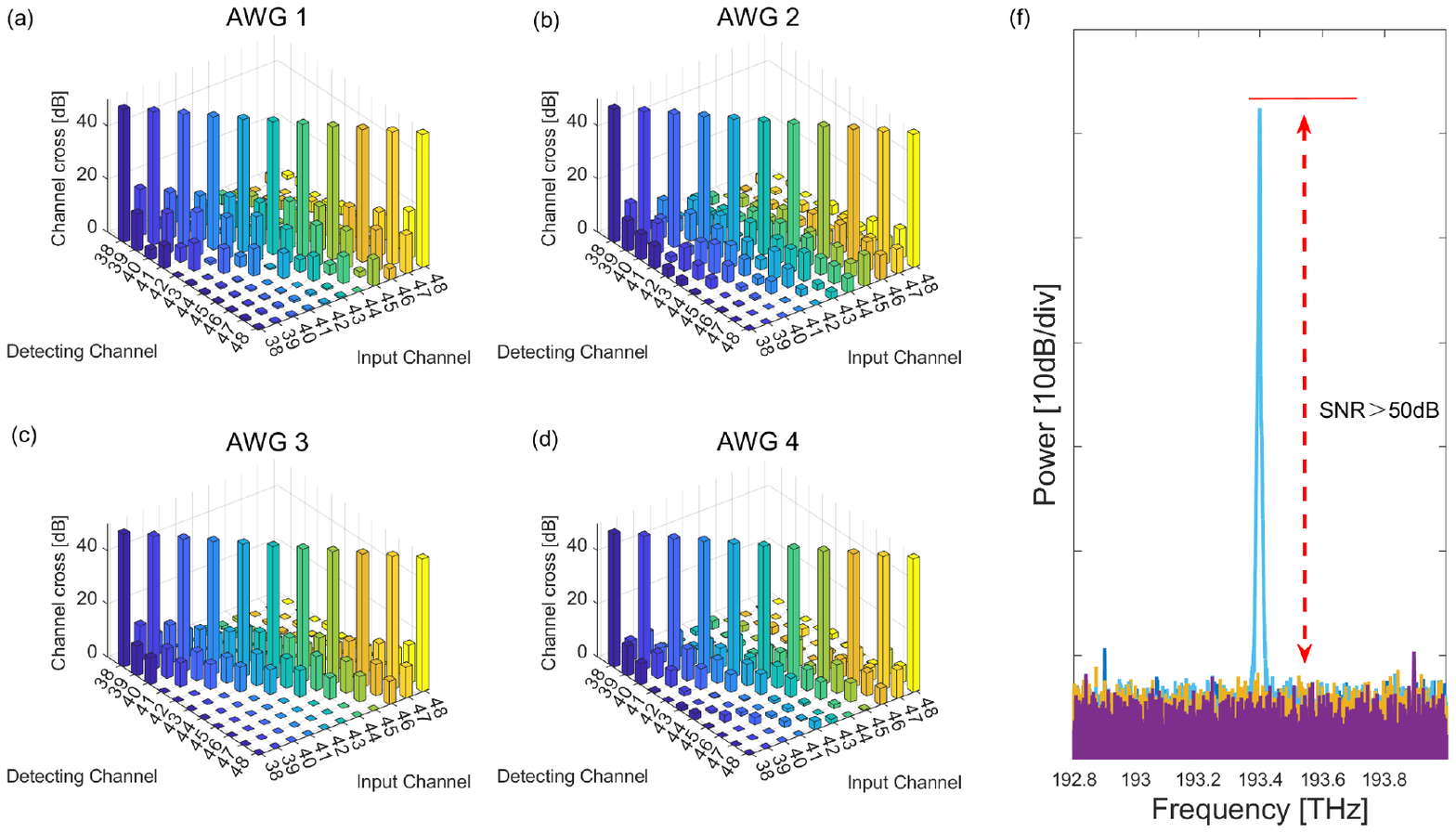}
		\caption{\label{Figs3} The measurement signal-to-noise ratio(SNR) of casecaded 32 channel arrayed waveguide grating filters with 100 GHz spacing. From \textbf{(a)} to \textbf{(d)} is the specific channel crosstalk from AWG number 1 to 4. \textbf{(f)}. The cascaded filtering effect can reach a signal-to-noise ratio of more than 50dB.} 
	\end{center}
\end{figure*}

In our system, the high signal-to-noise ratio (SNR) signal light is mainly accomplished by post-filtering with 32-channel waveguide fiber arrays. In the 'Monitor' part of the experimental setup (see Fig. ~\ref{Fig1}), the high-power pump light is first separated from the generated DKS comb lines by a 200 $GHz$ bandwidth fiber WDM, and then most of the power passes though the AWG by a 1:9 fiber beam splitter. The selected CH38 to CH47 comb lines are transported to another laboratory via a long fiber. Due to the low power of a single comb line, an auxiliary light (central wavelength in the CH21 channel) with a power of about 10 dbm is needed to lock the IM working point, so cascade of AWGs are used to get a pure signal light. We characterized the signal crosstalk of adjacent channels of a single AWG as shown in the Fig. ~\ref{Figs3}: the signal light is punched in from the input channel, and the output power is scanned with an optical power meter from CH38 to CH47. The obtained isolation of each adjacent channel of a single AWG can be kept above 28 dB, and the next nearest neighbor can be kept above 35 dB. The signal isolation of the adjacent channels of the cascaded AWGs is characterized to be above 50 dB by a optical spectrometer analyzer.

\subsection{The results of 10 HOM interference visibilities}\label{sec:four}

\begin{figure*}[htbp]
	\begin{center}
		\includegraphics[width=0.7\textwidth]{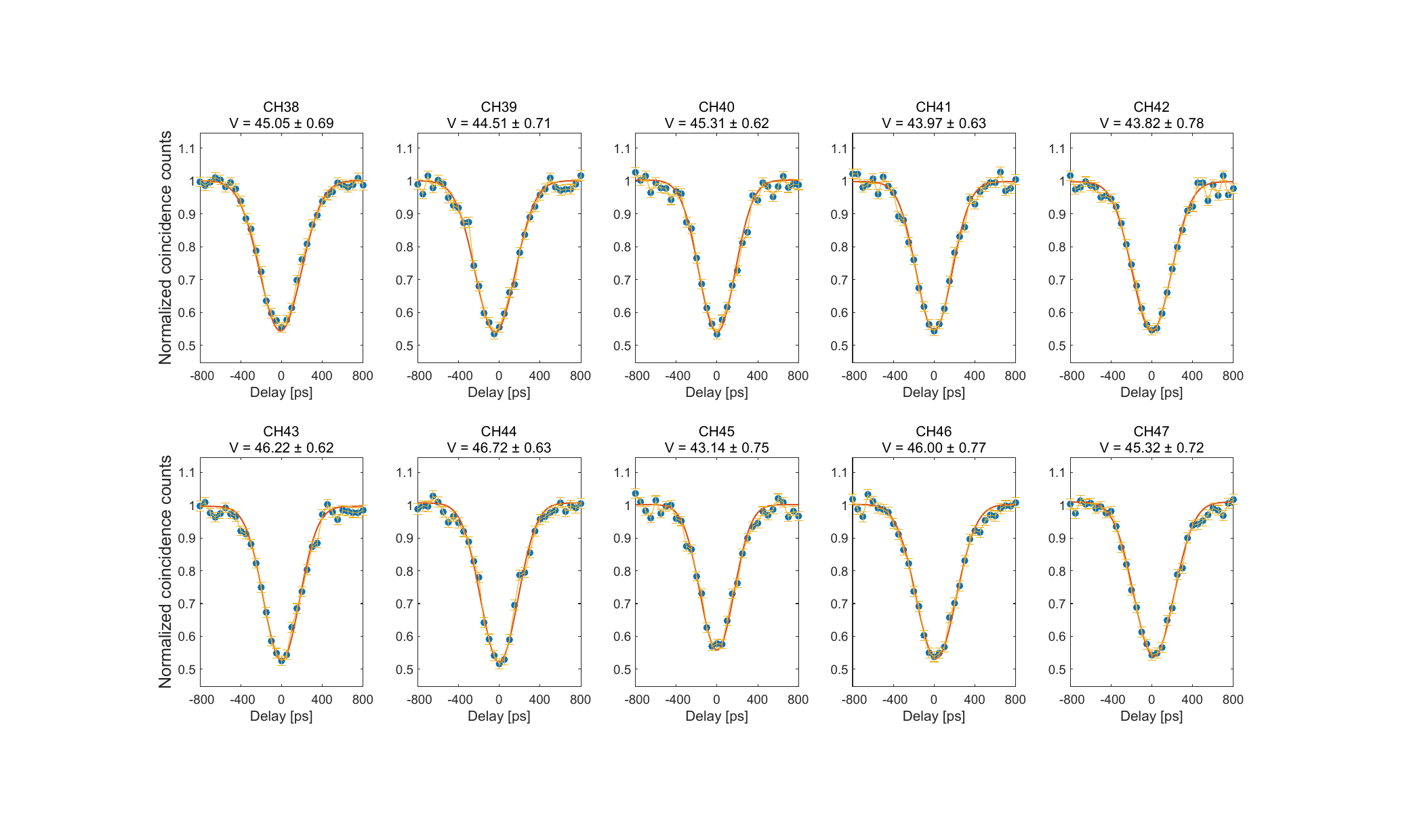}
		\caption{\label{Figs4}HOM interference visibility details of 10 comb-line pairs. The visibility distributes in the range of $43.14\%$ to $46.72\%$ which is measured for five times.} 
	\end{center}
\end{figure*}

As shown in Fig. ~\ref{Fig1} HOM interference part, the polarization has been controlled by the combination of PBS and PM-BS. The average photon number from different DKS comb lines is set to about 400,000 per second and the measurement interference results are shown in Fig.~\ref{Figs4}.

\clearpage
\end{document}